\documentclass[transaction]{IEEEtran}
\usepackage{graphicx}
\usepackage{epstopdf}
\usepackage[latin1]{inputenc}
\usepackage{amsmath,amssymb,amscd,latexsym,dsfont}
\usepackage[english]{babel}
\usepackage{tabularx,cite}
\usepackage{graphicx}
\usepackage{algpseudocode}
\usepackage{algorithm}
\usepackage{algorithmicx}
\usepackage{float}
\usepackage{multicol}
\usepackage{psfrag}
\usepackage{comment,psfrag,subfigure,enumerate}
\usepackage{color}
\usepackage{amsthm}

\ifCLASSINFOpdf
\else
\fi
\begin{document}
%
% paper title
% can use linebreaks \\ within to get better formatting as desired
\title{Finite Block-length Analysis of the Incremental Redundancy HARQ}

% author names and affiliations
% use a multiple column layout for up to three different
% affiliations
\author{
    \IEEEauthorblockN{Behrooz Makki, Tommy Svensson, Michele Zorzi, \emph{Fellow, IEEE}}
    \thanks{Behrooz Makki and Tommy Svensson are with the Department of Signals and Systems, Chalmers University of Technology, Gothenburg, Sweden, Email: \{behrooz.makki, tommy.svensson\}@chalmers.se}
    \thanks{Michele Zorzi is with the Department of Information Engineering, University of Padova, Padova, Italy, Email: zorzi@dei.unipd.it}
    \thanks{This work was supported in part by the Swedish Governmental Agency for Innovation Systems (VINNOVA) within the VINN Excellence Center Chase.}
}
\maketitle
%\onecolumn

\begin{abstract}
This letter studies the power-limited throughput of a communication system utilizing incremental redundancy (INR) hybrid automatic repeat request (HARQ). We use some recent results on the achievable rates of finite-length codes to analyze the system performance. With codewords of finite length, we derive closed-form expressions for the outage probabilities of INR HARQ and study the throughput in the cases with variable-length coding. Moreover, we evaluate the effect of feedback delay on the throughput and derive sufficient conditions for the usefulness of the HARQ protocols, in terms of power-limited throughput. The results show that, for a large range of HARQ feedback delays, the throughput is increased by finite-length coding INR HARQ, if the sub-codeword lengths are properly adapted.
\end{abstract}
%Then, compared to open-loop communication setups, the implementation of power-adaptive ARQ reduces the average power by ? and ? dB, if a maximum of 2 and 3 retransmissions is utilized, respectively.
% IEEEtran.cls defaults to using nonbold math in the Abstract.
% This preserves the distinction between vectors and scalars. However,
% if the conference you are submitting to favors bold math in the abstract,
% then you can use LaTeX's standard command \boldmath at the very start
% of the abstract to achieve this. Many IEEE journals/conferences frown on
% math in the abstract anyway.

% no keywords

% For peer review papers, you can put extra information on the cover
% page as needed:
% \ifCLASSOPTIONpeerreview
% \begin{center} \bfseries EDICS Category: 3-BBND \end{center}
% \fi
%
% For peerreview papers, this IEEEtran command inserts a page break and
% creates the second title. It will be ignored for other modes.
\IEEEpeerreviewmaketitle
\vspace{-2mm}
\section{Introduction}
%\textcolor{red}{kheili mohem: tu taghrib zadan low snr tu kaghaza eshtebah kardam. az makhraj chizi ke mimune $\sqrt{2gp}$ hastesh va $\sqrt{g^2P^2}$}
Hybrid automatic repeat request (HARQ) techniques are commonly used in wireless networks to combat the loss of data packets due to channel fading \cite{6477555,5684186,mimoarqkhodemun,5336856}.
%In HARQ schemes, if the receiver fails in decoding the data correctly, it asks for a retransmission and the data is retransmitted until it is correctly decoded or the maximum permitted number of retransmissions is reached \cite{6477555,536913,throughputdef,tuninetti2011,5336856,noisyARQkhodemun,1661837,outageHARQ}.
%is a well-established approach for wireless networks
%From an information-theoretic point of view, the HARQ systems can be viewed as channels with sequential feedback where, utilizing both forward error correction and error detection, the system performance is improved by retransmitting the data which has experienced poor channel conditions.
The performance of HARQ protocols is addressed in various papers, e.g., \cite{6477555,5684186,mimoarqkhodemun,5336856}, where the results are obtained under the assumption of asymptotically long codewords. On the other hand, in many applications, such as vehicle-to-vehicle and vehicle-to-infrastructure communications for traffic efficiency/safety or real-time video processing for augmented reality, the codewords are required to be short (in the order of $\sim 100$ channel uses) \cite{4657278,metisdel}. Thus, it is interesting to investigate the performance of HARQ protocols in the presence of finite-length codewords \cite{ARQrahul,ARQzorzikhodemun}.

In this letter, we study the data transmission efficiency of HARQ protocols utilizing codewords of finite length. The problem is cast as the maximization of the power-limited throughput in the presence of incremental redundancy (INR) HARQ feedback. The contributions of the paper are two-fold. 1) We use the recent results on the achievable rates of finite block-length codes \cite{5452208,6620483,yanginft} to analyze the throughput. With codewords of finite length, we derive closed-form expressions for the outage probabilities of the INR HARQ in different retransmission rounds and evaluate the effect of variable-length coding on the throughput. 2) We investigate the effect of feedback delay on the throughput. Particularly, we present sufficient conditions for the usefulness of HARQ protocols such that the use of HARQ increases the throughput compared to the open-loop communication setups. For a large range of HARQ feedback delays, the results show that the implementation of finite-length INR HARQ leads to throughput improvements.

%We investigate the effect of the codeword length on the optimal power allocation and the outage probability of the ARQ protocols. In particular, we show that, for codewords of length $L\ge 50$ channel uses, the performance of ARQ protocols is (almost) insensitive to the length of the codewords, in the sense that the changes in the outage probability are negligible for different codeword lengths. As demonstrated, considerable power efficiency improvement is achieved by the implementation of power-adaptive ARQ. For instance, consider Rayleigh fading channels, codewords of rate $1$ nats-per-channel-use (npcu) and target outage probability $10^{-3}.$ Then, compared to the open-loop communication setup, implementation of ARQ with a maximum of 2 and 3 transmissions reduces the average power by $17$ and $23$ dB, respectively, a result which is (almost) independent of the codewords length. With a maximum of $M=2$ transmissions, we derive closed-form solutions for the optimal, in terms of power-limited outage probability, power allocation between the ARQ transmissions. Finally, it is shown that with a maximum of $M=2$ transmissions the diversity gain of the ARQ protocol increases from 2 to 3, if optimal power allocation is utilized.

\vspace{-0mm}
\section{System model}
Consider a point-to-point communication setup following
\vspace{-0mm}
\begin{equation}
\vspace{-0mm}
Y = \sqrt{P}h X + Z,
\vspace{-0mm}
\end{equation}
where $P$ is the transmit power, $X$ is the unit-variance input message, $h$ denotes the fading coefficient and $Z \sim \mathcal{CN}(0,1)$ is the independent and identically distributed (iid) complex Gaussian noise added at the receiver. We consider an INR HARQ protocol in which each data packet is sent using a maximum of $M$ transmissions.

The system performance is studied in quasi-static conditions, e.g., \cite{mimoarqkhodemun,5336856}, where the channel coefficients remain constant during a packet transmission, and then change to other values according to the fading probability density function (pdf). As a motivation for this model, consider the LTE standard; as discussed in \cite{6477555}, each HARQ round corresponds to one transmission slot which is 0.5 milliseconds in the LTE standard. Also, for systems operating at a carrier frequency around 2.5 GHz and in the case that the receiver is moving with a speed of 2 km/h the coherence time is equal to 200 ms \cite{6477555}. This coherence time is 400 times larger than the time slot duration in LTE. Thus, with proper selection of the maximum number of
retransmissions, the quasi-static case can properly model the channel characteristics. (For more motivations for the quasi-static models, see \cite{5336856,mimoarqkhodemun}.)

Let us define the \emph{channel gain} as $g\doteq|h|^2.$ The results are given for Rayleigh fading channels where $h \sim \mathcal{CN}(0,1)$ and, as a result, $f_g(x)=e^{-x}$ with $f_g$ denoting the channel gain pdf. In each slot, the channel coefficient is assumed to be known by the receiver, which is an acceptable assumption in quasi-static conditions \cite{6477555,5684186,5452208,6620483,yanginft,5336856,mimoarqkhodemun}. However, no instantaneous channel state information is assumed to be available at the transmitter except the HARQ feedback bits.

%We consider Type-I ARQ with a maximum of $M-1$ retransmissions, i.e., the data is transmitted a maximum of $M$ times, and in each round the receiver disregards the previous messages, if received in error. Also, we define a packet as the transmission of a codeword along with all its possible retransmissions. Finally, the results are obtained for a frequency-hopping based scheme where the fading coefficient changes in each transmission independently.

\vspace{-0mm}
\section{Analytical results}
In this section, we study the throughput of the INR HARQ protocol utilizing codewords of finite length. We first obtain a closed-form expression for the throughput, following the same procedure as in \cite{5684186,5336856,mimoarqkhodemun}, except that we add the effect of feedback delay into the analysis. The throughput, given in (\ref{eq:etaasli}), is a function of a set of probabilities that depend on the sub-codewords' length. Hence, we use the results of finite-length codes \cite{5452208,6620483,yanginft} to obtain the probability terms. To find the probabilities, we need to use approximation and bounding techniques, as stated in Lemmas 1 and 2. Finally, we derive bounds for the desired HARQ feedback delays such that the implementation of HARQ increases the throughput compared to the open-loop communication setup. The details are presented as follows.

Using INR HARQ, $K$ information nats are encoded into a \emph{parent} codeword of length $l_{(M)}=\sum_{m=1}^M{l_m}$ channel uses. Then, the parent codeword is divided into $M$ sub-codewords of length $l_m, m=1,\ldots, M,$ channel uses which are sent in the successive transmission rounds. Thus, the equivalent data rate at the end of round $m$ is $R_{(m)}=\frac{K}{l_{(m)}},l_{(m)}=\sum_{n=1}^m{l_n},R_{(0)}\doteq\infty.$ In each round, the receiver combines all received sub-codewords to decode the message. The retransmission continues until the message is correctly decoded or the maximum permitted transmission round is reached.

Let $D$ (in channel uses) denote the HARQ feedback delay in each round. If the data transmission is stopped at the end of the $m$-th round, the total number of channel uses is
\begin{align}\label{eq:feedbacktm}
\tau_{(m)}=\left\{\begin{matrix}
l_{(m)}+mD,\,\,\,\,\,\,\,\,\,\,\,\,\,\,\,\,\,\,\text{       if  } m\ne M\\
l_{(M)}+(M-1)D,\,\,\,\,\text{if  } m= M
\end{matrix}\right.
\end{align}
which is based on the fact that in each retransmission round, except the last round, an acknowledgement/negative acknowledgement (ACK/NACK) signal is fed back to the transmitter. In this way, with some manipulations, the expected number of channel uses in each packet transmission period is found as
\begin{align}\label{eq:expectedtmfinite}
\mathcal{T}=\sum_{m=1}^M{l_m\Omega_{m-1}}+D\sum_{m=1}^{M-1}{\Omega_{m-1}},
\end{align}
where $\Omega_m$ denotes the probability that the message is not correctly decoded up to the end of the $m$-th round and $\Omega_0\doteq 1.$

If the message is correctly decoded in any round, all $K$ information nats are received by the receiver. Thus, following the same arguments as in \cite{5684186,5336856,mimoarqkhodemun}, the expected number of received information nats in each packet is given by
\begin{align}\label{eq:feedbackKm}
\mathcal{K}=K(1-\Pr(\text{Outage}))=K(1-\Omega_M),
\end{align}
where $\Pr(\text{Outage})=\Omega_M$ is the packet outage probability. Using (\ref{eq:expectedtmfinite}), (\ref{eq:feedbackKm}), the \emph{renewal-reward theorem} \cite{5684186,5336856,mimoarqkhodemun} and  $l_m=\frac{K}{R_{(m)}}-\frac{K}{R_{(m-1)}},$ the throughput (in nats-per-channel-use (npcu)) is obtained as
\begin{align}\label{eq:etaasli}
\eta=\frac{\mathcal{K}}{\mathcal{T}}=\frac{1-\Omega_M}{\sum_{m=1}^M{(\frac{1}{R_{(m)}}-\frac{1}{R_{(m-1)}})\Omega_{m-1}}+\frac{D^\text{f}}{R_{(M)}}\sum_{m=1}^{M-1}{\Omega_{m-1}}}.
\end{align}
Here, we have defined $D^\text{f}\doteq\frac{D}{l_{(M)}}$ which is referred to as the \emph{relative delay}, with respect to the maximum packet length.

To study the power-limited throughput of the INR protocol the final step is to calculate the probabilities $\Omega_m,m=1,\ldots,M.$
In the following, we use the recent results of \cite{5452208,6620483,yanginft} to find $\Omega_m$ for the cases with codewords of finite length. Let us first define an $(L,N,P,\delta)$ code as the collection of
\begin{itemize}
  \item An encoder $\Upsilon:\{1,\ldots,N\}\mapsto\mathcal{C}^L$ which maps the message $n\in\{1,\ldots,N\}$ into a length-$L$ codeword $x_n\in\{x_1,\ldots,x_N\}$ satisfying the power constraint
      \begin{align}\label{eq:code1}
\frac{1}{L}\left \| x_j \right \|^2\le P, \forall j.
\end{align}
  \item  A decoder $\Lambda:\mathcal{C}^L\mapsto\{1,\ldots,N\}$ satisfying the maximum error probability constraint
      \vspace{-2mm}
      \begin{align}\label{eq:code1}
\mathop {\max }\limits_{ \forall j}\Pr(\Lambda(y)\ne J|J=j)\le \delta
\end{align}
with $y$ denoting the channel output induced by the transmitted codeword according to $y.$
\end{itemize}
The maximum achievable rate of the code is defined as
\begin{align}\label{eq:achievablerateeq1}
 R_\text{max}(L,P,\delta)=\sup\left\{\frac{\log N}{L}:\exists (L,N,P,\delta)\, \text{code}\right\} \,\text{(npcu)}.
\end{align}
Considering quasi-static conditions, \cite{yanginft,6620483} have recently presented a very tight approximation for the maximum achievable rate (\ref{eq:achievablerateeq1}) as
\vspace{-2mm}
\begin{align}\label{eq:achievablerateeq2}
R_\text{max}(L,P,\delta)&= \sup\left\{R:\Pr(\log(1+gP)<R)<\delta\right\}\nonumber\\&-\mathcal{O}\left(\frac{\log L}{L}\right)\,\,\text{(npcu)},
\end{align}
which, for codes of rate $R$ npcu, leads to the following error probability \cite[eq. (59)]{yanginft}
\vspace{-2mm}
\begin{align}\label{eq:errorfiniteblock}
\delta(L,R,P)\approx E\bigg[Q\bigg(\frac{\sqrt{L}\left(\log(1+gP)-R\right)}{\sqrt{1-\frac{1}{(1+gP)^2}}}\bigg)\bigg].
\end{align}
Here, $U(x)=\mathcal{O}(V(x)),x\to\infty$ is defined as $\lim_{x\to\infty}\sup|\frac{U(x)}{V(x)}|<\infty $ and $E[.]$ is the expectation with respect to the channel gain $g.$ Also, $Q(.)$ denotes the Gaussian $Q$-function. Since the approximation (\ref{eq:errorfiniteblock}) has been shown to be very tight for sufficiently large values of $L$ \cite{5452208,6620483,yanginft}, for simplicity we will assume that it is exact in the following (especially when discussing bounds).
%According to \cite{yanginft,6620483}, the approximations in (\ref{eq:achievablerateeq2}) and (\ref{eq:errorfiniteblock}) are very tight for sufficiently large values of $L$. The choice of the minimum possible length depends on the required tightness of the approximations. Our analytical results hold for every value of $L$. For the simulation results, however, we consider the cases with $L\ge 100$ channel uses, for which the approximation is tight enough, and we do not consider shorter codewords. Our choice of $L\ge100$ as the minimum possible length is motivated by \cite[Fig. 2]{yanginft} where the relative difference of the exact and the approximate achievable rates is less than $2\%$ for $L\ge100.$  We are further strengthened in our choice of the sub-codewords length by reports such as \cite{4657278}, which suggest the practical codes of interest for, e.g., vehicle to vehicle communication to be in the range of 100-300 channel uses. In the meantime, although the approximation is not tight for small $L$'s and the results should not be fully trusted in that case, we observe the same qualitative conclusions as in the case of $L\ge 100,$ when the simulations are run for very short (practically not interesting) codewords.

From (\ref{eq:achievablerateeq2})-(\ref{eq:errorfiniteblock}), the probability that the data is not decodable in rounds $n=1,\ldots,m$, of the INR, i.e., $\Omega_m,$ is found as
\begin{align}\label{eq:probomegam}
\Omega_m=E\bigg[Q\bigg(\frac{\sqrt{l_{(m)}}\left(\log(1+gP)-R_{(m)}\right)}{\sqrt{1-\frac{1}{(1+gP)^2}}}\bigg)\bigg].
\end{align}
Here, (\ref{eq:probomegam}) is based on the fact that 1) with a quasi-static condition, the same fading realization $g$ is experienced in all rounds of a packet, 2) the receiver combines all received signals of a packet to decode the message and, 3) for a given value of $K$ nats, $W_{(m)}(x)=\frac{\sqrt{l_{(m)}}\left(\log(1+xP)-\frac{K}{l_{(m)}}\right)}{\sqrt{1-\frac{1}{(1+xP)^2}}}$ is an increasing function of $l_{(m)}$ and, thus, $A_m\subset A_n,n<m$ for quasi-static channels, where $A_m$ is the event that the data is not decoded in rounds $1,\ldots,m.$

For Rayleigh fading conditions, $\Omega_m$ is found as
\vspace{-1mm}
\begin{align}\label{eq:omegam1}
\Omega_m=\int_0^\infty{e^{-x}Q\left(\frac{\sqrt{l_{(m)}}(\log(1+Px)-R_{(m)})}{\sqrt{1-\frac{1}{(1+Px)^2}}}\right)\text{d}x}
\end{align}
which does not have a closed-form expression. Lemmas 1-2 approximate/bound the probabilities $\Omega_m,\forall m,$ as follows.

\emph{\textbf{Lemma 1.}} The probabilities $\Omega_m,\forall m,$ are approximated by
\vspace{-1mm}
\begin{align}
\Omega_m=\left\{\begin{matrix}
\frac{1}{2}\bigg(1-e^{\frac{1}{P}}\sum_{i=0}^\infty{\frac{1}{i!}(\frac{-e^{R_{(m)}}}{P})^i}\Big(1-\text{erf}(\frac{K-i-1}{\sqrt{2l_{(m)}}})\Big)\,\,\,\,\,\,\,\,\,\,\,\,\,\,\,\,\,\,\,\,\,\,\,\,\,\,\,\,\,\,\,\,\,\,\,\,\,\,\,\,\,\, \\ -\text{erf}(-\frac{R_{(m)}\sqrt{l_{(m)}}}{\sqrt{2}})\bigg),\,\,\,\,\,\,\,\,\,\,\,\,\,\,\,\,\,\,\,\,\,\,\, \text{For high SNRs}
\\
1-\frac{b_m}{\sqrt{2\pi}}e^{-\theta_m}\left(e^{\sqrt{\frac{\pi}{2b_m^2}}}-e^{-\sqrt{\frac{\pi}{2b_m^2}}}\right),\,\,\,\,\,\,\text{For all SNRs}\,\,\,\,\,\,\,\,\,\,\,\,\,\,\,\,\,\,\,\,\,\,\,\,\,\,\,
\end{matrix}\right.\nonumber
\end{align}
where $\text{erf}(.)$ represents the error function, $\theta_m\doteq\frac{e^{R_{(m)}}-1}{P}$ and $b_m\doteq\sqrt{\frac{l_{(m)}P^2}{e^{2R_{(m)}}-1}}$.
\begin{proof}
At medium/high signal-to-noise ratios (SNRs), (\ref{eq:probomegam}) is approximated by $\Omega_m=$ $\int_0^\infty e^{-x}Q({\sqrt{l_{(m)}}(\log(1+xP)-R_{(m)})})\text{d}x$ which leads to
\vspace{-3mm}
\begin{align}\label{eq:omegam2}
\Omega_m&\mathop  = \limits^{(a)} \sqrt{\frac{l_{(m)}}{2\pi}}\int_{e^{-{R_{(m)}}}}^\infty{\frac{1-e^{-\frac{te^{R_{(m)}}-1}{P}}}{t}e^{-\frac{l_{(m)}}{2}(\log t)^2}\text{d}t}\nonumber\\&\mathop  = \limits^{(b)}\sqrt{\frac{l_{(m)}}{2\pi}}\int_{e^{-{R_{(m)}}}}^\infty{t^{-1}e^{-\frac{l_{(m)}}{2}(\log t)^2}\text{d}t}-\nonumber\\&
\sqrt{\frac{l_{(m)}}{2\pi}}e^\frac{1}{P}\sum_{i=0}^\infty{\frac{1}{i!}(\frac{-e^{R_{(m)}}}{P})^i\int_{e^{-R_{(m)}}}^\infty{t^{i-1}e^{-\frac{l_{(m)}}{2}(\log t)^2}\text{d}t}}
\nonumber\\&={\frac{1}{2}}\Big(1-\text{erf}(-\frac{R_{(m)}\sqrt{l_{(m)}}}{\sqrt{2}})-\nonumber\\&\,\,\,\,\,\,\,\,\,\,\,\,\,\,\,\,\,\,\,\,e^{\frac{1}{P}}\sum_{i=0}^\infty{\frac{1}{i!}(\frac{-e^{R_{(m)}}}{P})^i}\Big(1-\text{erf}(\frac{K-i-1}{\sqrt{2l_{(m)}}})\Big)\Big).
\end{align}
Here, $(a)$ is obtained by partial integration, variable transform $t=(1+Px)e^{-R_{(m)}}$ and the definition of the Gaussian $Q$-function with $Q(\infty)=0,$ $\frac{\text{d}Q(s(x))}{\text{d}x}=\frac{-1}{\sqrt{2\pi}}\frac{\text{d}s(x)}{\text{d}x}e^{-\frac{s(x)^2}{2}}$. Then, $(b)$ follows from the Taylor expansion of the first exponential term, and the last equality is obtained by some manipulations and the definition of the error function $\text{erf}(x)=\frac{2}{\sqrt{\pi}}\int_0^x{e^{-t^2}}\text{d}t.$

For the second approximation approach, we implement
$Q\bigg(\frac{\sqrt{l_{(m)}}\left(\log(1+xP)-R_{(m)}\right)}{\sqrt{1-\frac{1}{(1+xP)^2}}}\bigg)\simeq Z_m$ with
\begin{align}\label{eq:linearZm}
Z_m(x)\mathop  = \limits^{(c)}\left\{\begin{matrix}
1 & x\le \theta_m-\sqrt{\frac{\pi}{2b_m^2}}\\
\frac{1}{2}-\frac{b_m}{\sqrt{2\pi}}(x-\theta_m) & x\in\big[\theta_m-\sqrt{\frac{\pi}{2b_m^2}},\theta_m+\sqrt{\frac{\pi}{2b_m^2}}\big]\\
0 & x\ge \theta_m+\sqrt{\frac{\pi}{2b_m^2}}
\end{matrix}\right.
\end{align}
which results in
\begin{align}\label{eq:lowsnromegam}
&\Omega_m=\int_0^{\infty}{e^{-x}Z_m(x)\text{d}x}\mathop  = \limits^{(d)}1-\frac{b_m}{\sqrt{2\pi}}e^{-\theta_m}\left(e^{\sqrt{\frac{\pi}{2b_m^2}}}-e^{-\sqrt{\frac{\pi}{2b_m^2}}}\right).
\end{align}
%\begin{align}
%&\Omega_m=\int_0^{\theta_m-\sqrt{\frac{\pi}{2b_m^2}}}{e^{-x}\text{d}x}+\int_{\theta_m+\sqrt{\frac{\pi}{2b_m^2}}}^{\theta_m-\sqrt{\frac{\pi}{2b_m^2}}}{(\frac{1}{2}+\frac{b_m\theta_m}{\sqrt{2\pi}}-\frac{b_m}{\sqrt{2\pi}}x)e^{-x}\text{d}x}
%\end{align}
%\vspace{-1mm}
%\begin{align}\label{eq:lowsnromegam}
%&\Omega_m=\int_0^{\frac{R_{(m)}}{P}}{e^{-x}Q\Big(-\frac{\sqrt{l_{m}}}{\sqrt{2Px}}\Big)\text{d}x}+\int_{\frac{R_{(m)}}{P}}^\infty{e^{-x}Q\Big(\frac{\sqrt{l_{m}}}{\sqrt{2Px}}\Big)\text{d}x}\nonumber\\&\mathop  = \limits^{(c)}1-e^{-\frac{R_{(m)}}{P}}-\int_0^\frac{R_m}{P}{e^{-x}\Big(\frac{1}{2}+\frac{1}{\sqrt{2\pi}}\sum_{n=1}^\infty{a_{2n}(\frac{l_{(m)}}{2P})^{n}x^{-n}}\Big){}\text{d}x}\nonumber\\&+\int_\frac{R_m}{P}^\infty{e^{-x}\Big(\frac{1}{2}+\frac{1}{\sqrt{2\pi}}\sum_{n=1}^\infty{a_{2n}\big(\frac{l_{(m)}}{2P}\big)^{n}x^{-n}}\Big){}\text{d}x}\nonumber\\&\mathop  = \limits^{(d)}\frac{1}{2}+\frac{1}{\sqrt{2\pi}}\sum_{n=1}^\infty{a_{2n}\big(\frac{l_{(m)}}{2P}\big)^{n}\Big(\Gamma(1-n)-2\Gamma\big(1-n,\frac{R_{(m)}}{P}\big)\Big)}.
%\end{align}
%$C_{(m)}=\frac{-\sqrt{l_{(m)}}R_{(m)}}{\sqrt{2P}}$
Here, $(c)$ is obtained by using the linearization technique for the function $Q\bigg(\frac{\sqrt{l_{(m)}}\left(\log(1+xP)-R_{(m)}\right)}{\sqrt{1-\frac{1}{(1+xP)^2}}}\bigg)$ at $x=\theta_m$ and $(d)$ follows from (\ref{eq:linearZm}) and some manipulations.
\end{proof}
\emph{\textbf{Lemma 2.}} The probabilities $\Omega_m,\forall m,$ are bounded by
\begin{align}\label{eq:lemmabound}
\begin{array}{l}
\,\,\,\,\,\,\,\,\,\,\,\,\,\,\,\,\,\,\,\,\,\,\,\,\,\,\,\,\,\,\,\,\,\,\,v_m\le \Omega_m\le u_m,\forall m,\epsilon>0,\\
v_m=\frac{1}{2}\big(1-\text{erf}(\frac{-\theta_mb_m}{\sqrt{2}})-e^{\frac{1-2\theta_mb_m^2}{2b_m^2}}(1-\text{erf}(\frac{1-b_m^2\theta_m}{\sqrt{2}b_m}))\big), \\
u_m=1-\frac{e^{-\theta_m}+e^{-\psi_m}}{2}+\frac{1}{2}e^{\alpha_m}P^{-\epsilon l_{(m)}}\Gamma(1-\epsilon l_{(m)},\psi_m+\frac{1}{P}),\\
\psi_m\doteq\frac{e^{(R_{(m)}+\frac{\epsilon}{2})}-1}{P},\alpha_m\doteq\frac{1}{P}+K\epsilon+\frac{l_{(m)}\epsilon^2}{2}.
\end{array}
\end{align}
%where $\Gamma(.,.)$ is the incomplete Gamma function.
\begin{proof}
$Q(x)$ is a decreasing function and $W_{(m)}(x)=\frac{\sqrt{l_{(m)}}\left(\log(1+xP)-R_{(m)}\right)}{\sqrt{1-\frac{1}{(1+xP)^2}}}$ is concave in $x$. Thus, from (\ref{eq:probomegam}), a lower bound on $\Omega_m$ is obtained if $W_{(m)}(x)$ is replaced by its first-order Taylor expansion at any point. Using the Taylor expansion of $W_{(m)}(x)$ at $x=\theta_m,$ we have
\vspace{-3mm}
\begin{align}\label{eq:prooftheoremx}
\Omega_m&\ge\int_0^\infty{e^{-x}Q(b_m(x-\theta_m))\text{d}x}\nonumber\\&\mathop  = \limits^{(e)}\int_0^\infty{\frac{b_m(1-e^{-x})}{\sqrt{2\pi}}e^{-\frac{b_m^2}{2}(x-\theta_m)^2}\text{d}x}=v_m,
\end{align}
where $(e)$ comes from partial integration and $v_m$ given in (\ref{eq:lemmabound}) is found by some manipulations and the definition of error function.

The upper bound is found by
\vspace{-2mm}
\begin{align}\label{eq:equpperbound}
\begin{array}{l}
\Omega_m\mathop  \le \limits^{(f)} 1-e^{-\theta_m}+\frac{1}{2}\int\limits_{\theta_m}^\infty{e^{-x}e^{-\frac{l_{(m)}}{2}(\log(1+Px)-\frac{K}{l_{(m)}})^2}\text{d}x}\\\,\,\,\,\,\,\,\,\,\,\mathop  \le \limits^{(g)} 1-\frac{e^{-\theta_m}+e^{-\psi_m}}{2}+\\\,\,\,\,\,\,\,\,\,\,\,\,\,\,\,\,\,\,\,\,\,\,\frac{e^{K\epsilon+\frac{\epsilon^2l_{(m)}}{2}}}{2}\int_{\theta_m}^\infty{(1+Px)^{-\epsilon l_{(m)}}e^{-x}\text{d}x}=u_m.
\end{array}
\end{align}
Here, $(f)$ is obtained by (\ref{eq:omegam1}), using the inequality
\vspace{-1mm}
\begin{align}
Q(x)\le\left\{\begin{matrix}
1,\,\,\,\,\,\,\,\,\,\,\,\,\,\,\,\,\, \text{if}\,\,x<0\\
\frac{1}{2}e^{-\frac{x^2}{2}},\,\,\,\, \text{if}\,\,x\ge0
\end{matrix}\right.
\end{align}
and removing the denominator in (\ref{eq:omegam1}) when $x\ge \theta_m.$ Then, $(g)$ follows from $(a-b)^2\ge \max\{0,2\epsilon a-2b\epsilon-\epsilon^2\},\forall a\ge b,\epsilon>0,$ and some manipulations. Finally, the last equality is given by the definition of the incomplete Gamma function $\Gamma(n,x)=\int_x^\infty{t^{n-1}e^{-t}\text{d}t}$.
\end{proof}
Depending on the SNR/the sub-codewords length, the approximations/bounds in Lemmas 1-2 are useful in different conditions. Finally, to enjoy the benefits of the INR HARQ the channel code should satisfy the following requirements: 1) A parent code that can be punctured into rate-optimized sub-codewords and 2) a decoder with performance close to (\ref{eq:probomegam}) for all retransmissions. There exist several practical finite-length code designs, e.g., \cite{refldpcinr1,refldpcinr2}, that satisfy these requirements.
%Shown in Fig.1 is the ...
\vspace{-4mm}
\subsection{On the Effect of Feedback Cost}
With no HARQ (open-loop setup), the parent codeword of length $l_{(M)}$ is sent in \emph{one shot.} Thus, with uniform power allocation, on which we focus, the outage probability is $\Omega_M$, the same as in the HARQ-based scheme. Also, as there is no feedback, the throughput of the non-HARQ scheme is found as
\vspace{-1mm}
\begin{align}\label{eq:etanoharq}
\eta^\text{Open-loop}=\frac{K}{l_{(M)}}(1-\Pr(\text{Outage}))=R_{(M)}(1-\Omega_M).
\end{align}

From (\ref{eq:etaasli}), (\ref{eq:etanoharq}), the intuition behind the HARQ protocols is as follows. If the channel quality is low, all possible transmissions of the HARQ are used and the system performance will be the same as in the case with no HARQ (except for the additional feedback delays). But, if the channel quality is high, the message may be correctly decoded at the end of the $m$-th, $m<M$ round, and the channel uses for rounds $m+1,\ldots,M$ are saved. The cost of this \emph{gambling} is the cost for feedback, i.e., the term $D\sum_{m=1}^{M-1}{\Omega_{m-1}}$ in (\ref{eq:expectedtmfinite}). Hence, depending on the feedback delay  and the channel conditions, using HARQ may or may not improve the throughput.
%Note that with asymptotically long codeword assumption, the feedback delay can be ignored, as it is in \cite{6477555,throughputdef,tuninetti2011,mimoarqkhodemun,5336856,noisyARQkhodemun,1661837,outageHARQ,powerarq2007}. However, the feedback delay should be taken into account when considering finite-length codes.

To find the acceptable range of feedback delays, we can maximize (\ref{eq:etanoharq}) for a given power and then sweep on different values of relative feedback delay, i.e., $D^\text{f}$ in (\ref{eq:etaasli}), to find the maximum value of $D^\text{f}$ for which the HARQ-based approach leads to higher throughput, compared to non-HARQ scheme. Sufficient conditions for the usefulness of the HARQ, i.e., lower bounds on the acceptable range of feedback delays such that the throughput is improved by the HARQ, are obtained as follows.

For every given value of power and information nats, optimize $l_{(M)}$ in terms of the open-loop throughput (\ref{eq:etanoharq}). Consider the same packet length $l_{(M)}$ and fixed-length coding, i.e., $l_m=\frac{l_{(M)}}{M},\forall m,$ for HARQ, which is not necessarily optimal for HARQ-based scheme, in terms of throughput. Then, as the HARQ and the non-HARQ schemes have the same outage probability, the throughput is increased by HARQ if it results in less expected delay. Hence, from (\ref{eq:expectedtmfinite}), a sufficient condition for the usefulness of HARQ is given by $\mathcal{T}\le l_{(M)}$ leading to $D^\text{f}\le r$ with
\vspace{-2mm}
\begin{align}\label{eq:boundasli}
r=\frac{1-\frac{1}{M}\sum_{m=1}^M{\Omega_{m-1}}}{\sum_{m=1}^{M-1}{\Omega_{m-1}}}.
\end{align}
Using Lemma 2, $\Omega_0=1$ and because (\ref{eq:boundasli}) is a decreasing function of $\Omega_m,\forall m,$ we have
\vspace{-1mm}
\begin{align}\label{eq:bounddelay}
 \frac{M-1-\sum_{m=1}^{M-1}{u_m}}{M(1+\sum_{m=1}^{M-2}{u_m})}\le r\le \frac{M-1-\sum_{m=1}^{M-1}{v_m}}{M(1+\sum_{m=1}^{M-2}{v_m})}
\end{align}
with $u_m$ and $v_m$ derived in (\ref{eq:lemmabound}). Note that, while (\ref{eq:boundasli}) provides sufficient conditions for the HARQ feedback delay with fixed-length coding, variable-length HARQ indeed does better, and larger range of feedback delays are tolerated in the variable-length coding scheme.
The numerical results are presented in the sequel.
\vspace{-7mm}
\section{Numerical Results and Conclusions}
According to \cite{yanginft,6620483}, the approximations in (\ref{eq:achievablerateeq2}) and (\ref{eq:errorfiniteblock}) are very tight for sufficiently long sub-codewords, and the tightness increases with the sub-codewords' length.
%Our analytical results hold for every value of $L$ in (\ref{eq:errorfiniteblock}).
For the numerical results, we consider the cases with $l_m\ge 100,\forall m,$ channel uses, for which the approximation is tight enough, and we do not consider shorter sub-codewords. Our choice of $l_m\ge100$ as the minimum possible length is motivated by \cite[Fig. 2]{yanginft} where the relative difference of the exact and the approximate achievable rates is less than $2\%$ for the cases with codewords of length $\ge100.$ We are further motivated for our choice of the sub-codewords length by reports such as \cite{4657278}, which suggest the practical codes of interest for, e.g., vehicle-to-vehicle communication to be in the range of $100-300$ channel uses.
%In the meantime, although the approximation is not tight for small $L$'s and the results should not be fully trusted in that case, we observe the same qualitative conclusions as in the case of $L\ge 100,$ when the simulations are run for very short (practically not interesting) codewords.

Shown in Fig. 1a is the throughput achieved by the variable- and the fixed-length coding INR HARQ and $D^\text{f}=0$. Here, the results are obtained for the cases with a maximum of $M=2$ transmissions, both the number of information nats and the sub-codewords' length are optimized in terms of throughput, and the simulation results are compared with the ones obtained via the approximation techniques of Lemma 1. Setting $K=600$ nats, Fig. 1b demonstrates the throughput gain of the variable-length HARQ, i.e., $\Delta=\frac{\eta-\eta^\text{Open-loop}}{\eta^\text{Open-loop}}\%$ with $\eta$ being the throughput in variable-length HARQ. Finally, using fixed-length coding, Fig. 1c shows the acceptable range of feedback delay and compares the results with (\ref{eq:boundasli}) and the bounds developed in (\ref{eq:bounddelay}), i.e., when the probabilities are bounded via Lemma 2. In harmony with Lemma 1, the approximations in (\ref{eq:omegam2}) and (\ref{eq:lowsnromegam}) are tight at high and all SNRs, respectively (Fig. 1a). Also, variable-length coding INR leads to throughput increment, compared to the fixed-length coding INR and the open-loop setup, especially at high SNRs (Figs. 1a-1b).
%Optimizing the throughput in variable-length coding scheme, we have $l_1\ge l_2$ (\textcolor{blue}{Although} not demonstrated in the figures, we observe the same order $l_n\ge l_m,n<m,$ for the cases with $M=3,4$ as well.).
As illustrated in Fig. 1c, the bounds developed in Lemma 2 and the sufficient condition in (\ref{eq:boundasli}) are very tight for a large range of SNR. Also, the acceptable range of HARQ feedback delay, in terms of throughput, is very low at medium SNRs while its effect is relaxed at low and high SNRs. Finally, the acceptable range of feedback delay decreases with $K$. To summarize, using INR HARQ with finite-length codes results in throughput increment for a large range of feedback delays, particularly when the SNR increases.
%\section*{Acknowledgement}
%This work was supported in part by the Swedish Governmental Agency for Innovation Systems (VINNOVA) within the VINN Excellence Center Chase.
\begin{figure}
\vspace{-3mm}
\centering
  % Requires \usepackage{graphicx}
  \includegraphics[width=0.98\columnwidth]{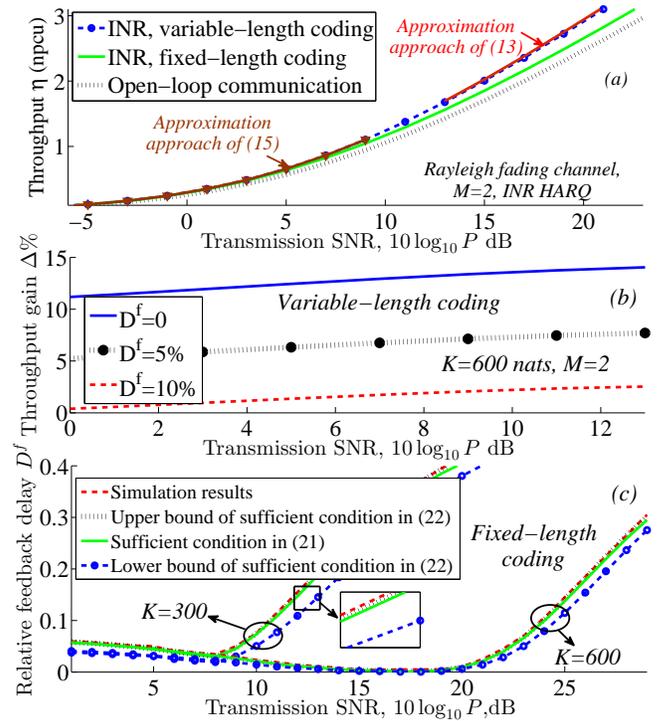}\\\vspace{-3mm}
\caption{(a): Throughput, (b): throughput gain and (c): acceptable relative feedback delay vs SNR, $M=2$. In subplot (a), both the number of information nats and the sub-codewords' length are optimized in terms of throughput. In subplots (b) and (c), $K=600$ and $K=300,600$ nats, respectively.}\label{figure111}
\vspace{-5mm}
\end{figure}
\vspace{-3mm}
\bibliographystyle{IEEEtran} %lic.bst is the style file
\bibliography{masterfiniteblock}
\vfill
% that's all folks
\end{document}